\documentclass{aa}  
\usepackage{graphicx}
\usepackage{txfonts}
\begin{document}

\title{Dynamics of Nearby Groups of Galaxies:\\
the role of the cosmological constant}
\author{S. Peirani\inst{1}
          \and
          J.A. de Freitas Pacheco\inst{2}
          }

   \institute{Institut d'Astrophysique de Paris, 98 bis Bd Arago, 75014 Paris, France - Unit\'e mixte de 
         recherche 7095 CNRS - Universit\'e Pierre et Marie Curie \\
              \email{peirani@iap.fr}
         \and
             Observatoire de la C\^ote d'Azur - Laboratoire Cassiop\'ee - UMR 6202 - BP 4229 - 06304 - Nice Cedex 4 - France\\
             \email{pacheco@oca.eu}
             }

   \date{\today}

 
  \abstract
  {Different cosmological data are consistent with an accelerated expansion produced
   by an exotic matter-energy component, dubbed ``dark-energy''. A cosmological constant
   is a possibility since it satisfies most of the observational constraints.
In this work, the consequences of such a component in the dynamics 
   of groups of galaxies is investigated, aiming to detect possible effects in
   scales of the order of few Mpc. 
The Lema\^itre-Tolman model was modified by the inclusion of the cosmological constant term and,
from the numerical solution of the equations of motion, a velocity-distance
   relation was obtained. This relation depends on two parameters: the central core mass and
the Hubble parameter. The non-linear fit of such a relation to available data
   permitted to obtain masses for five nearby groups of galaxies and for
   the Virgo cluster as well as estimates of the Hubble constant.
The analysis of the present results indicates that the velocity-distance relation derived from
the modified Lema\^itre-Tolman model as well as that derived from the ``canonical''  model give equally acceptable
   fits to the existent data. As a consequence, any robust conclusion on the effects of the cosmological constant
in the dynamics of groups could be established. The mean value of the Hubble parameter derived from the present
study of local flows is $H_0 = 65\pm 7$ km/s/Mpc.

\bigskip

{\bf key words.} Galaxies: kinematics and dynamics; Cosmological parameters;

}
{}{}{}{}


   \maketitle
%

\section{Introduction}

Small groups of galaxies are very common structures in the universe and may contribute
up to $\sim$ 50\% of its matter content (Huchra \& Geller 1982; Geller \& Huchra 1983;
Nolthenius \& White 1987). Early estimates of mass-to-light ($M/L$) ratios for
groups based on the virial relation lead to values typically of the order of
170~$M_{\odot}/L_{B,\odot}$ (Huchra \& Geller 1982). However, new and high quality data
on galaxies situated in nearby groups (Karachentsev 2005 and references therein) yield 
values around 10-30~$M_{\odot}/L_{B,\odot}$, considerably smaller than
past estimates. If these new estimates are correct, then the local matter density
would be only a fraction of the global matter density. 
The virial relation gives trustful masses only if groups are in dynamical equilibrium. Observers
assume, in general, that this is the case if the crossing time, ratio between a suitable defined
radius, characterizing the dimension of the group and the typical velocity of galaxies
 belonging to the system, is less than the
Hubble time. Using cosmological simulations, Niemi et al. (2007) identified groups with
the same algorithm used by observers. Their analysis indicates that about 20\% of them are not gravitationally bound
and that such a fraction increases when the apparent magnitude limit of the survey increases. Moreover,
Niemi et al. (2007) do not have found any correlation between the virial ratio $2T/\mid W\mid$
and the crossing time, a result independent of the magnitude limit of the mock catalogue and that
weakens the criterium usually adopted to characterize the dynamical equilibrium.

Besides the question related to the mass estimates of groups, a second and long-standing
problem concerns the fact that dispersion of the peculiar velocities over the Hubble flow
is quite small, usually referred as the ``coldness'' of the local velocity flow (Sandage
\& Tammann 1975; Giraud 1986; Schlegel et al. 1994; Ekholm et al. 2001). The presence
of the dark energy has been invoked as a possible explanation for the smoothness of
the local Hubble flow (Chernin 2001; Teerikorpi et al. 2005). 
Dark matter simulations by Governato et al. (1997) for cosmological models 
with $\Omega_m = 1$ or $\Omega_m = 0.3$ are, according to these
authors, unable to produce systems embedded in regions having ``cold'' flows, i.e., with
1-D dispersion velocities of the order of 40-50 km/s. From simulations based on a $\Lambda CDM$ cosmology,
Macci\`o et al. (2005) and Peirani \& de Freitas Pacheco (2006, hereafter PP06) estimated
values for the 1-D dispersion velocity, averaged within a sphere of $\sim$ 3 Mpc radius, of
80 and 73 km/s respectively.  New simulations
were recent reported by Hoffman et al. (2007), who have compared results issued from
$CDM$ and $\Lambda CDM$ cosmologies with identical parameters, except for the presence or not
of the cosmological constant term. They claim that no significant differences are noticed
in the velocity flow around galaxies having properties similar to those observed in the neighborhood
of the Milky Way (MW).

In the case of groups constituted by a central and dominant mass, corresponding to a single or 
to a pair of massive galaxies and an external ``cloud'' of low mass satellites, an alternative approach 
to the virial relation was already proposed in the eighties by Lynden-Bell (1981) and 
Sandage (1986), based on the Lema\^itre-Tolman (LT)
model (Lema\^itre 1933; Tolman 1934). The LT model describes quite well the dynamics of a bound
central core embedded by an extended halo, which approaches asymptotically a homogeneous
Friedmann background. In this situation, three main distinct regions can be distinguished: i) the
central core, in which the shell crossing has already occurred, leading to energy exchanges
which transform radial into random motions; ii) the zero-velocity surface, boundary which separates
infalling and expanding bound shells and iii) the ``marginally'' bound surface (zero total energy),
segregating bound and unbound shells. Density profiles resulting from the LT model were
discussed by Olson \& Silk (1979) and an application of this model to the velocity field close to the
Virgo cluster was made by Hoffman et al. (1980), Tully \& Shaya (1984) an Teerikorpi et al. (1992),
among others. The LT model, modified by the inclusion of a cosmological constant, was revisited by
PP06, who applied their results for a sample of galaxies in the outskirt of the Virgo cluster and 
for dwarf galaxies in the vicinity of the M31-MW pair.
 
The velocity-distance ($v$-$R$) relation is a photograph of the kinematic status of the different shells 
at a given instant. Such a relation, derived either from the ``canonical'' or the modified LT model, depends 
only on two parameters: the core mass and the Hubble parameter. Thus, if pairs of values (v, R) are known 
for satellite galaxies of a given group, a non-linear fit can be performed between data and the theoretical 
$v$-$R$ relation, permitting a simultaneous determination of the core/central galaxy mass and of the 
Hubble parameter (PP06). In this paper, the $v$-$R$ relation modified by the inclusion of a cosmological constant,
was calculated for the present age of the universe by using an improved integration algorithm with respect to
that employed originally by PP06. The fit of the resulting numerical values gives a slight different solution,
which was applied to data on the groups M81, Sculptor, CenA/M83 and IC342/Maffei-I, allowing estimates of 
their masses as well as of the Hubble parameter $H_0$. The previous analysis by PP06 of the Local Group and of the
Virgo cluster was also revisited. As we shall see, the inclusion or not of the cosmological constant in the 
LT model does not affect considerably the resulting masses. However, the resulting values of the Hubble 
parameter are systematically higher for the LT model when compared to the modified LT model, but
mean values resulting from both approaches are consistent, within the 
uncertainties, with other independent determinations. This paper is organized as follows: in Section 2 a short 
review of the model is presented; in Section 3 an application to nearby groups is made 
and, finally, in Section 4 the main conclusions are given.

\section{The Model}

The spherical collapse model of a density perturbation dates back to the work by Gunn \& Gott (1972), who
described how a small spherical patch decouples from the homogeneous expanding background, slows down,
turns around and collapses, forming finally a virialized system. The inclusion of other forms of energy
besides gravitation has been the subject of many investigations as, for instance, those by
Lahav et al. (1991), Wang \& Steinhardt (1998), Maor \& Lahav (2005) among others.

Here we follow the procedure adopted in our previous paper (PP06), since we intend to obtain, {\it for
the present age of the universe}, the velocity profile of small galaxies subjected to the gravitational field
of the central massive object. This approach supposes that satellites do not contribute significantly
to the total mass of the group, that orbits are mainly radial and that they do not form a relaxed
system. Effects of non-zero angular momentum orbits will be discussed later. 
We assume also that displacements
of the satellite galaxies, here associated to the outer shells, develop at redshifts when the formation
of the mass concentration around the core is nearly complete (see, for instance, Peebles 1990) or, in
other words, that any further mass accretion is neglected. Under
these conditions, the equation of motion for a spherical shell of mass $m$, moving radially in the
gravitational field created by a mass $M (>> m)$ inside a shell of radius $R$, including the cosmological
constant term is
\begin{equation}
\frac{d^2R}{dt^2}=-\frac{GM}{R^2}+\Omega_{\Lambda}H^2_0R\, .
\label{motion1}
\end{equation}
This equation has a first integral, which expresses the energy conservation, given by
\begin{equation}
\left(\frac{dR}{dt}\right)^2=\frac{2GM}{R}+\Omega_{\Lambda}H^2_0R^2+2E\, ,
\label{integral1}
\end{equation}
where $E$ is the energy per unit of mass of a given shell. Notice that eq.~\ref{integral1} is
similar to the Hubble equation in which the energy plays the role of the curvature constant.
In order to solve numerically eq.~\ref{motion1},
it is convenient to define the dimensionless variables: $y=R/R_0$, $x=tH_0$ and $u=\dot{R}/H_0R_0$, where
$R_0$ is the radius of the zero-velocity surface, i.,e., $\dot{R}(R_0)\equiv$ v$(R_0)=0$. In 
term of these variables, the equations above can be recast as
\begin{equation}
\frac{d^2y}{dx^2}=-\frac{A}{2y^2}+\Omega_{\Lambda}y
\label{motion2}
\end{equation}
and
\begin{equation}
u^2=\frac{A}{y}+\Omega_{\Lambda}y^2+K\, ,
\label{integral2}
\end{equation}
where we have introduced the constants $A=2GM/(H^2_0R^3_0)$ and $K=2E/(H_0R_0)^2$. In order
to integrate the equation of motion, initial conditions have to be imposed. We took for
the initial value of the shell radius $y_i=10^{-3}$, corresponding to an initial proper dimension of
about 1 kpc. When $y<<1$, the gravitational term dominates the right side of eq.~\ref{motion2}
and an approximate solution, valid for small values of the radius is
$y\simeq (9A/4)^{1/3}x^{2/3}$, from which the initial value of the dimensionless time
$x_i$, corresponding to the adopted value of $y_i$, can be estimated. Using the initial
value $y_i$, the initial value of the velocity $u_i$ can be estimated from eq.~\ref{integral2},
for a given value of $K$, i.e., of the shell energy. The constant $A$ is determined from
the integration of eq.~\ref{integral2} subjected to the conditions $u(y=1)=0$ and
\begin{equation}
x(y=1)=\int^{\infty}_0\frac{dz}{(1+z)\sqrt{\Omega_{\Lambda}+\Omega_m(1+z)^3}}
\end{equation}
which imply for
the zero-velocity shell an energy $K=-(A+\Omega_{\Lambda})$. We have obtained $A=3.7683$ for
$\Omega_{\Lambda}$ = 0.7, a value 3\% higher than that obtained previously by PP06. Using the definition
of $A$, the central mass can be estimated by the relation below, {\it if the radius of the
zero-velocity surface is known}, namely,
\begin{equation}
M=4.23\times 10^{12}h^2R_0^3 \,\,M_{\odot}\, ,
\label{mass}
\end{equation}
where $R_0$ is in Mpc and $h$ is the Hubble parameter in units of $100 km/s/Mpc$.  This 
result is essentially the basis 
of the method proposed by Lynden-Bell (1981) and Sandage (1986) to estimate the mass of 
the Local Group. The inclusion of the cosmological constant modifies the numerical factor and,
as emphasized by PP06, masses derived by this procedure are, for the same $R_0$, about 30\% higher than 
those derived neglecting the effet of the cosmological constant.

Once the value of $A$ is known (notice that the value of $A$ varies according to the considered
age of the universe), eq~\ref{motion2} can be integrated for different
values of $K$ or, equivalently, for different values of the initial velocity.
The inclusion of the cosmological constant modifies the general picture of the LT model. The central
core in which shell crossing has already occurred and the zero-velocity surface are still present.
However, for bound shells ($K<0$) which will reach the zero-velocity surface in the future, the
turnaround occurs only if $K<K_c = -4.06347$, where $K_c$ corresponds to the energy for which
the maximum expansion radius coincides with the zero-gravity surface (ZGS). For energy values higher than
the critical value $K_c$, once the shell crosses the ZGS (located at $y_c=1.391$), the
acceleration is positive and there is no fallback. This behavior is illustrated in figure 1,
where the evolution of shells having different energies $K$ is shown. The shell with $K=-6.3$
reached the maximum expansion at $\sim$ 8.0 Gyr ago and has already collapsed. The shell
with $K=-5.0$ attained the maximum expansion at $\sim$ 3.8 Gyr ago and it is still
collapsing. Galaxies identified presently with such a shell have negative velocities.
The shell with the particular energy $K=-4.4683$ has just reached the maximum expansion
or the zero-velocity surface and will collapse completely within $\sim$ 13.8 Gyr
from now. Finally, for the shell with $K=-4.06$ the zero-velocity radius is just beyond
the ZGS and the collapse will never occur.

\begin{figure}
\rotatebox{0}{\includegraphics[width=\columnwidth]{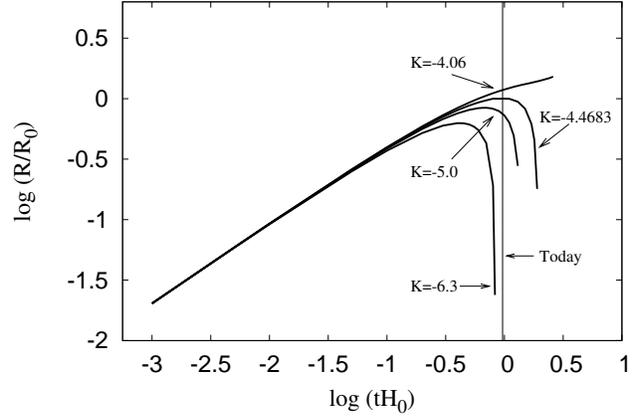}}
\caption{Evolution of shells with different energies K.}
 \label{fig1}
 \end{figure}

A fit of our numerical solution gives for the present velocity-distance relation
\begin{equation}
v(R)=-\frac{0.976H_0}{R^n}\left(\frac{GM}{H^2_0}\right)^{(n+1)/3} + 1.377H_0R
\label{velocity1}
\end{equation}
with $n=0.627$. Using the definition and the value of $A$, it is trivial to show that the equation
above satisfies the condition v$(R_0)=0$.

This equation was tested by PP06 on a group of
galaxies issued from cosmological simulations and whose properties were similar to those of
Local Group. The fit of eq~\ref{velocity1} to simulated data permitted to recover quite confidently
the mass of the central pair of galaxies.

\subsection{Angular momentum effects}

If orbits are not purely radial, the equation of motion (eq.~\ref{motion1}) becomes
\begin{equation}
\frac{d^2R}{dt^2}=-\frac{GM}{R^2}+\Omega_{\Lambda}H^2_0R+\frac{j^2}{R^3}\, ,
\end{equation}
where $j$ is the specific angular momentum of the shell. Numerical simulations indicate that
the specific angular momentum is well represented by a power law, e.g., $j=\kappa R^{\alpha}$,
with $\alpha=1.1\pm 0.3$ (Bullock et al. 2001). The coefficient $\kappa$ varies with time, reflecting
the halo mass accretion history. Here, in order to study the angular momentum effects, we assume
$\alpha = 1$ and take $\kappa$ as a constant. Under these conditions, in dimensionless variables,
the equation of motion can be written as
\begin{equation}
\frac{d^2y}{dx^2}=-\frac{A}{2y^2}+\Omega_{\Lambda}y+\frac{K_J}{y}\, ,
\end{equation}
where $K_J=(\kappa/H_0R_0)^2$. The equation above was integrated by using the same procedure
as before, for $K_J$=0.1 and $K_J$=1.0. If the dimensionless velocity profile is fitted as 
previously by the relation $u = -b/y^n + by$, the resulting parameters for the different values
of $K_J$ are given in table 1. For comparison, the corresponding values for the constant $A$ and
for the zero-gravity surface $y_{ZG}$ are also given. 

\begin{table}
\caption[]{Angular momentum effects on constants and fitting parameters}
\begin{flushleft}
\begin{tabular}{lllll}
\hline
$K_J$&b&n&A&$y_{ZG}$\\
\hline
0.0&1.377&0.627&3.7683&1.391\\
0.1&1.319&0.690&3.9516&1.379\\
1.0&1.156&0.900&5.6353&1.296\\
\hline
\end{tabular}
\end{flushleft}
\end{table}
 
A simple analysis of table 1 indicates that, as expected, the inclusion of the
``centrifugal" force term steepens the velocity profile and decreases
the radius of the zero-gravity surface. Moreover, the required value of the 
constant $A$ increases for higher values of the angular momentum and, as
a consequence, for a given $R_0$ the derived masses are still more important
than those derived from the modified or the  ``canonical" TL model.

\section{Application to nearby groups}

In the past years, a large amount of data on nearby groups have been obtained by different observers, in
particular by Karachentsev and collaborators. New dwarf galaxies have been discovered as a 
consequence of searches on the POSS II and ESO/SERC plates (Karachentseva and Karachentsev 1998, 2000)
as well as on ``blind'' HI surveys (Kilborn et al. 2002). Distances to individual members of nearby
groups have been derived from magnitudes of the tip of the red giant branch (Karachentsev 2005 and
references therein), which have permitted a better membership assignment and a more trustfull dynamical
analysis.

In order to apply eq.\ref{velocity1}, distances and velocities of
 each galaxy with respect to the center 
of mass should be computed. The radial distance $R$ is simply given by
\begin{equation}
R^2=D^2+D_g^2-2~D~D_g~cos~\theta\, ,
\end{equation}
where $D$ is the distance to the center of mass, $D_g$ is the distance to the considered galaxy and $\theta$
is the angular separation between the galaxy and the center of mass. Observed velocities are
generally given in the heliocentric system and were converted into the Local Group rest frame by using the 
prescriptions of the RC2 calalog. If $V$ and $V_g$ are respectively the velocities of the center of mass and
of the galaxy with respect to the LG rest frame, the velocity difference along the radial direction between
both objects is
\begin{equation}
V(R)=V_g~cos~\alpha-V~cos~\beta\, ,
\end{equation}
where $\beta=\theta+\alpha$ and $tan~\alpha=D~sin~\theta/(D_g-D~cos~\theta)$. The final list of
galaxies constituting each group excludes objects with uncertain distances and/or velocities as well
as objects beyond the zero energy surface, which are supposed to be unbound. The latter are chosen
after a first analysis in which the zero energy radius is roughly
estimated for each group.
Finally, for all considered galaxies, we affected an error equal to 10\% of
the respective values to both their
radial distances $R$ and velocities $V$. It is worth mentioning that increasing this
latter value up to 20\% doesn't affect significantly the estimations of $H_0$
and the mass M of each studied group.

\subsection{The M81 group}

Karachentsev et al. (2002a) presented a detailed study of the M81 complex. A distance of 3.5 Mpc was 
estimated from the brightness of the tip of the red giant branch, based on
HST/WFPC2 images of different members of the association. Using distances and radial velocities
of about 50 galaxies in and around the M81 complex, Karachentsev et al. (2002a) estimated
the radius of the zero-velocity surface as $R_0=(1.05\pm 0.07)$ Mpc and, using the LT model, they
derived a total mass within $R_0$ equal to $(1.6\pm 0.3)\times 10^{12} M_{\odot}$. Karachentsev
\& Kashibadze (2006) found a slightly lower value for the radius of the zero-velocity surface
around the pair M81/M82, i.e., $R_0=(0.89\pm 0.05)$ Mpc, corresponding to a total mass inside $R_0$ of
$(1.03\pm 0.17)\times 10^{12} M_{\odot}$.

As explained in PP06, our analysis follows a different approach. We performed a non-linear fit of
eq.~\ref{velocity1} to the available data, searching for the best values of the mass inside $R_0$ 
and of the Hubble constant which minimize the scatter. This procedure for the M81 complex gives
a total mass of $(9.2\pm 2.4)\times 10^{11} M_{\odot}$, which is consistent with the revised
value by Karachentsev \& Kashibadze (2005). The derived value of the Hubble constant (in
units of 100~km/s/Mpc) is
$h = 0.67\pm 0.04$. We emphasize that the quoted errors are estimates based on the spread of
values derived from the fitting procedure and not formal statistical errors. Figure 2a shows
the velocity-distance relation based on these data. Solid points are actual data and the solid 
curve is the best fit of eq.~\ref{velocity1}.

\subsection{The CenA/M83 complex}

Direct imaging of dwarf galaxies in the Centaurus A (NGC 5128) group were obtained by 
Karachentsev et al. (2002b, 2007). They have shown that these galaxies are concentrated essentially
around Cen A and M83 (NGC 5236) and that their distances to the Local Group are 3.8 Mpc and 4.8 Mpc
respectively. Using velocities and distances of individual members, the radius of the zero-velocity
surface around Cen A was estimated to be $R_0 = (1.44\pm 0.13)$ Mpc, leading to a total mass inside $R_0$ of
$(6.4\pm 1.8)\times 10^{12} M_{\odot}$. According to the authors, effects of the cosmological constant
were taken into account. Woodley (2006) has also performed a dynamical investigation of the 
Cen A group and, based on different mass indicators, he estimated for Cen A group a 
mass of $(9.2\pm 3.0)\times 10^{12} M_{\odot}$.

Figure 2b shows the velocity-distance diagram based on the available data. The solid line represents
again the best fit of eq.~\ref{velocity1}. Our analysis gives for Cen A a mass of 
$(2.1\pm 0.5)\times 10^{12} M_{\odot}$, which is a factor 3-4 lower than the aforementioned estimates. We
will discuss later the consequences of these mass determinations. The resulting value of the Hubble
parameter issued from the fitting procedure is $h = 0.57\pm 0.04$.

\subsection{The IC342/Maffei-I group}

A recent investigation on these groups was performed by Karachentsev et al. (2003a). They
found that seven dwarf galaxies are associated to IC342 group, at an average distance of 3.3 Mpc
from the Local Group. The Maffei-I association consists of about eight galaxies, with an uncertain 
distance estimate of about 3 Mpc. According to Karachentsev et al. (2003a), the total mass 
of this complex inside the zero-velocity surface $R_0=(0.9\pm 0.1)$ Mpc is 
$(1.07\pm 0.33)\times 10^{12} M_{\odot}$, a value which agrees with virial estimates, according
to those authors.

\begin{figure}
\rotatebox{0}{\includegraphics[width=8.cm]{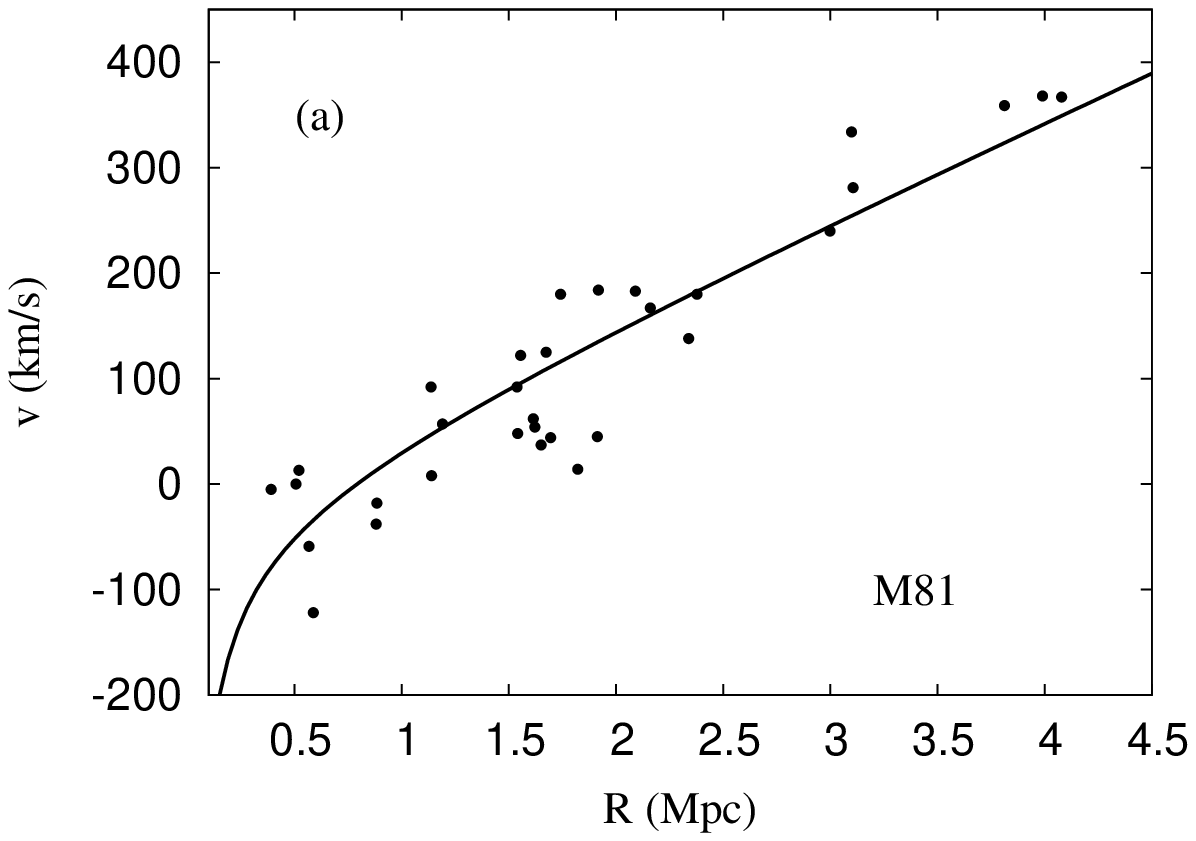}}
\rotatebox{0}{\includegraphics[width=8.cm]{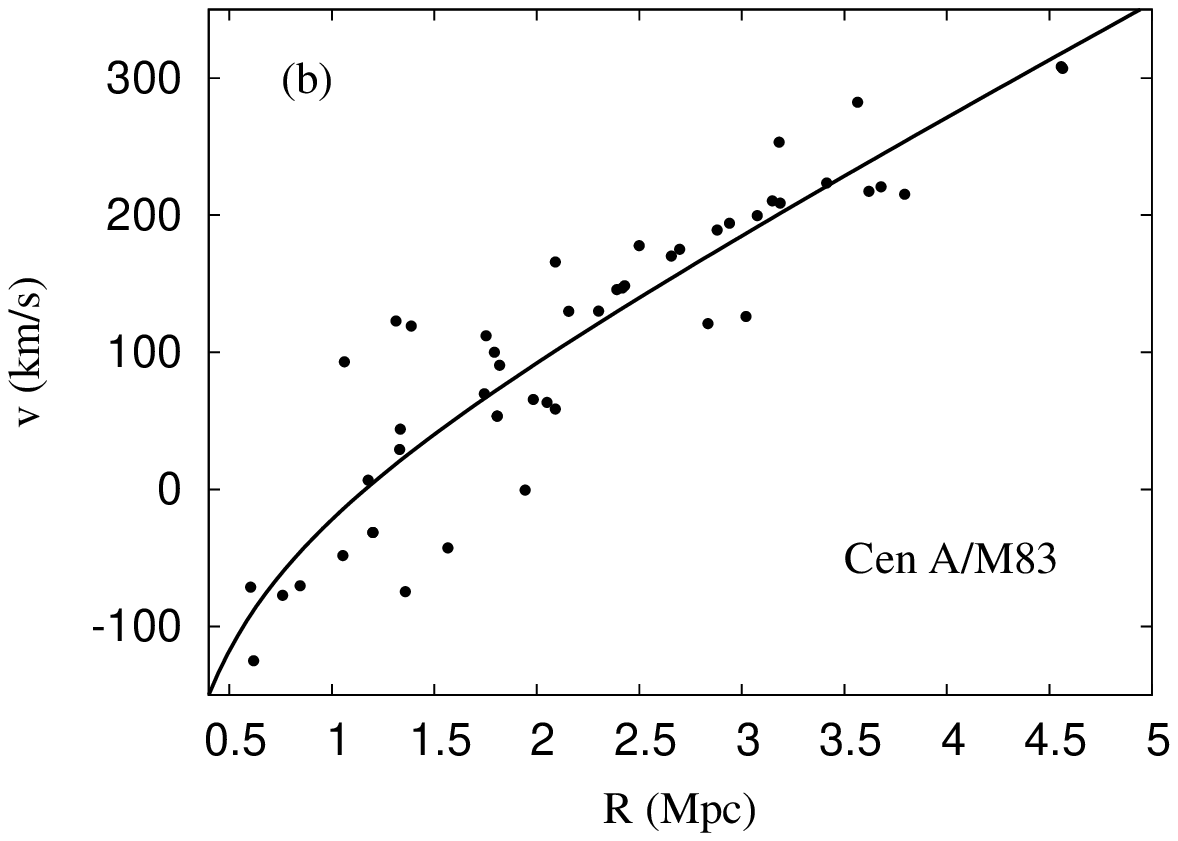}}
\rotatebox{0}{\includegraphics[width=8.cm]{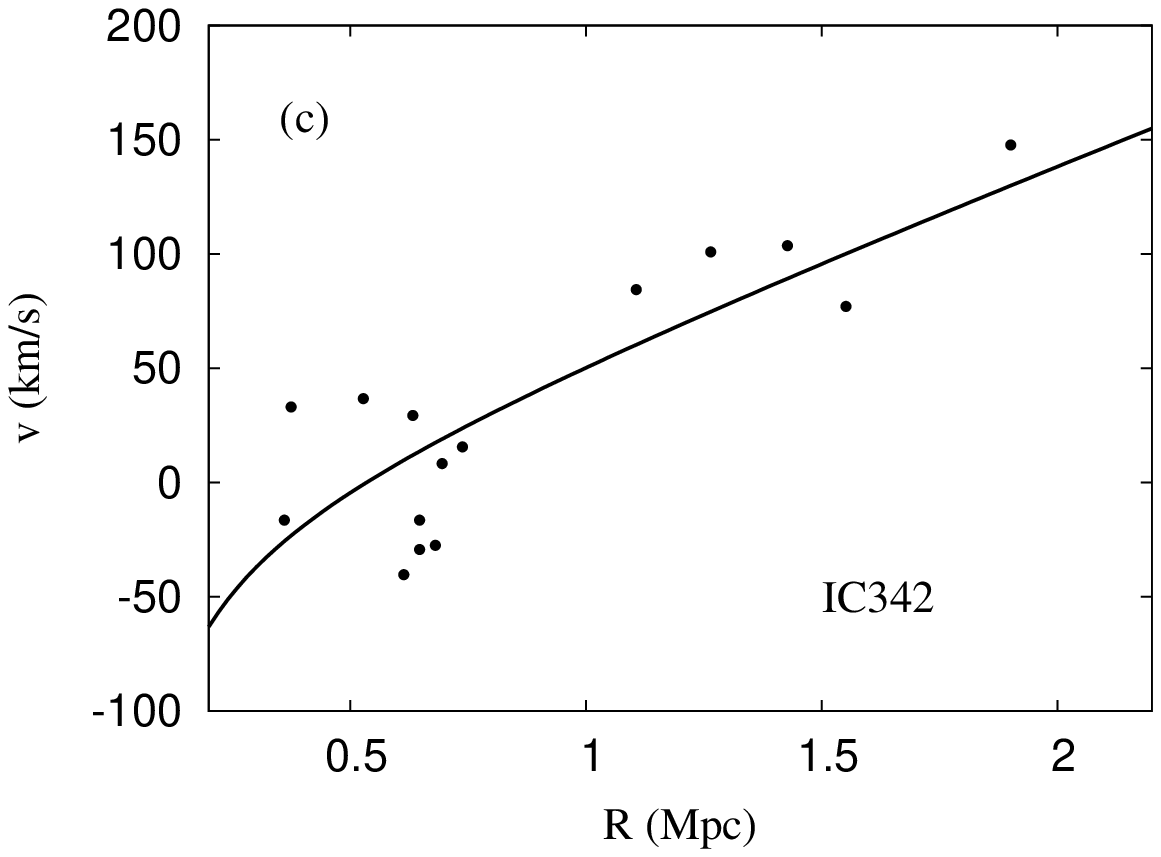}}
\rotatebox{0}{\includegraphics[width=8.cm]{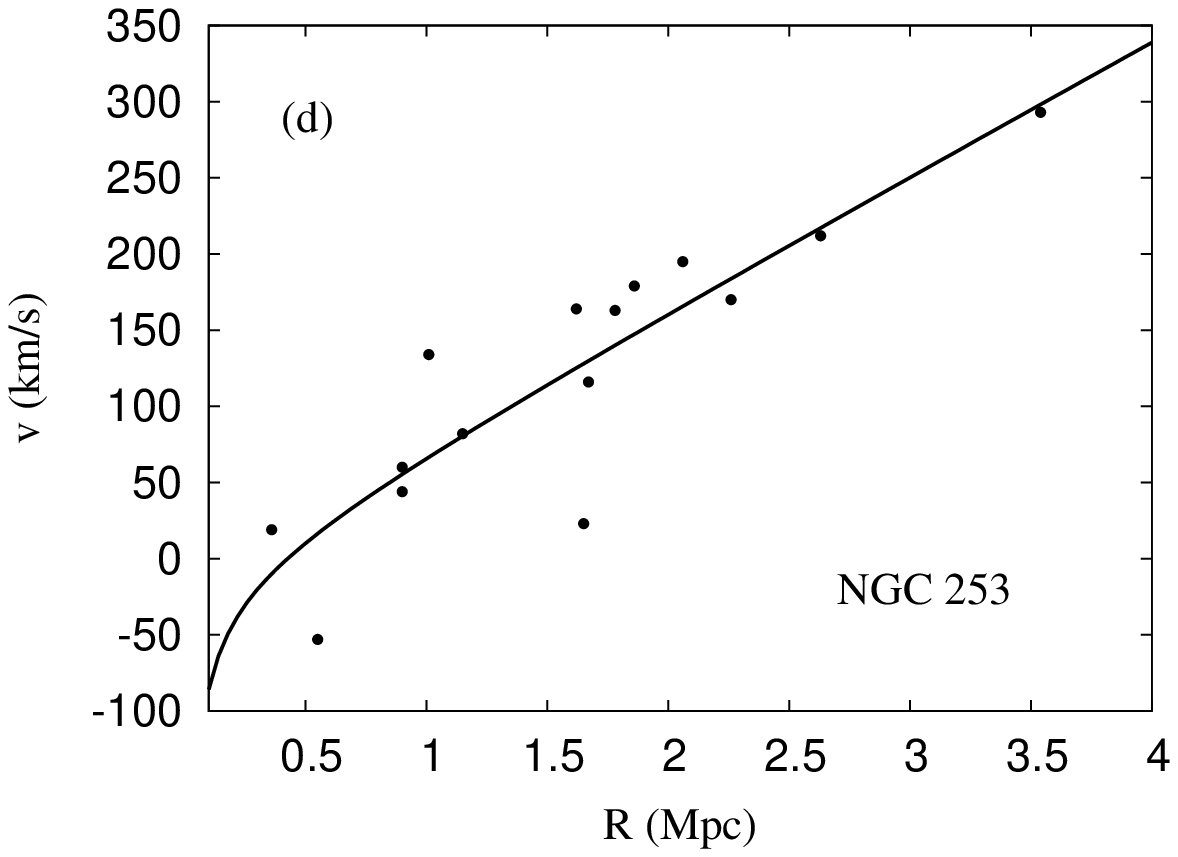}}
\caption{ Velocity-distance diagrams based on available data relative
 to (a) the M81 group, (b) the CenA/M83 complex, (c) the IC342/Maffei-I
 group and (d) the NGC 253 group. Solid lines are best fit to eq.~\ref{velocity1}.}
 \label{fig1}
 \end{figure}

Our own analysis of the same data leads to a total mass which is about a factor of 5 smaller, namely,
$(2.0\pm 1.3)\times 10^{11} M_{\odot}$. The best fit of the velocity-distance relationship (solid
curve) to data (solid points) is shown in figure 2c. Our results suggest that the zero-velocity
surface has a radius $R_0\simeq 0.53$ Mpc and can hardly be as high as the value given by Karachentsev
et al. (2003a), as a simple inspection of our plot indicates. The simultaneous estimate of the Hubble
constant from these data gives $h = 0.57\pm 0.10$.

\subsection{The NGC 253 (Sculptor) group}

This association was studied by Karachentsev et al. (2003b), who described the system as
small, loose concentrations of galaxies around NGC300, NGC253 and NGC7793. The authors
estimated the zero-velocity radius as been $R_0=0.7\pm 0.1$ Mpc and a total mass of
$(5.5\pm 2.2)\times 10^{11} M_{\odot}$. 

The non-linear fit of eq.~\ref{velocity1} to the data (figure 2d) gives a total mass of
$(1.3\pm 1.8)\times 10^{11} M_{\odot}$ and the large uncertainty suggests that data is
probably incomplete, in particular for objects with negative velocities (falling into the core).
The derived Hubble constant is $h = 0.63\pm 0.06$, whose determination is less affected by
the absence of galaxies with negative velocities.

\begin{table}
\caption[]{Properties of Groups: masses are in units of $10^{12} M_{\odot}$
and mass-to-light ratios are in solar
units for the B-band. Columns 2-3 correspond to the modified LT model, while
the last column gives masses derived for the ``canonical'' LT model.}
\begin{flushleft}
\begin{tabular}{llll}
\hline
Group&Mass $(\Omega_{\Lambda}=0.7)$& M/L &Mass $(\Omega_{\Lambda}=0)$\\
\hline
M31/MW &$2.4\pm 0.8$&$60\pm 20$&$2.5\pm 0.8$\\
M81 &$0.92\pm 0.24$&$56\pm 20$&$1.3\pm 0.4$\\
NGC 253&$0.13\pm 0.18$&$9\pm 8$&$0.12\pm 0.18$\\
IC 342&$0.20\pm 0.13$&$10\pm 8$&$0.22\pm 0.16$\\
CenA/M83&$2.1\pm 0.5$&$51\pm 20$&$2.2\pm 0.6$\\
Virgo&$1400\pm 300$&$580\pm 106$&$1800\pm 400$\\ 
\hline
\end{tabular}
\end{flushleft}
\end{table}

We have also revisited our previous analysis of the Virgo cluster. In order to be 
consistent with the hypothesis of our model, only galaxies with virgocentric distances
higher than 3.5 Mpc were selected, since most of the mass of the cluster is contained
inside a sphere of $\sim 7^o$ radius ($\sim$ 2.2 Mpc). The total luminosity of the cluster, 
$L=2.4\times 10^{12}~L_{B,\odot}$, was taken from Sandage, Bingelli \& Tammann (1985).

\subsection{Masses and $M/L$ ratios}

Table 2 summarizes our mass estimates, including revised values for the Local Group and the Virgo
cluster derived in our previous work (PP06). Mass-to-light ratios were computed by using photometric 
data of NED (http://nedwww.ipac.caltech.edu) and
LEDA (http://leda.univ-lyon1.fr) databasis. Asymptotic magnitudes for a given object in both 
databasis are sometimes quite discordant. For instance, in the case of IC 342, NED gives B=11.24 while 
LEDA gives B=9.10. This is an extreme case, but differences in the range 0.3-0.5 mag are present for the other 
objects. Face to these uncertainties, we have simply adopted the average B-luminosity derived from 
asymptotic magnitudes and reddening given in both databasis as well as distances mentioned above.
Notice that luminosities derived by such a procedure are, on the average, smaller by a factor of two
than those given for instance by Karachentsev (2005), which explains essentially the differences 
between his $M/L$ ratio values and the present estimates. Thus, uncertainties in the M/L ratios reflect uncertainties in
the mass determination, asymptotic magnitudes and distances. Simple inspection of these numbers indicates 
that the $M/L$ ratio increases for more massive systems, a well known behavior.

\subsection{Effects of the cosmological constant}

How the cosmological constant affects the results? The original $M$-$R_0$ relation
derived from the LT model is
\begin{equation}
M=\frac{\pi^2}{8G}\frac{R_0^3}{T_0^2}\, ,
\label{mass2}
\end{equation}
where $T_0$ is the age of the universe. For a flat $\Lambda CDM$ cosmology, $T_0=g(\Omega)/H_0$ where
\begin{equation}
g(\Omega)=\int^{\infty}_0\frac{dz}{(1+z)\sqrt{\Omega_{\Lambda}+\Omega_m(1+z)^3}}\, .
\end{equation}
Adopting $\Omega_m$=0.3, one obtains $g(\Omega)=0.964$ and, replacing the resulting age in eq.~\ref{mass2}, one
obtains a numerical coefficient about 30\% smaller than of eq.~\ref{mass}, as PP06 have already emphasized.
Thus, in the ``canonical'' LT model the cosmological constant affects only the age determination
and, for a given value of $R_0$, the resulting masses are smaller by that factor in comparison with
masses derived from eq.~\ref{mass}.

Another approach, adopted in PP06 and in the present work, consists to fit the theoretical velocity-distance
relation to data and derive consistently the total mass and the Hubble parameter. In this case, to analyze the
effects of the cosmological constant we have to compare the predictions for both models. The $v$-$R$ relation
for the ``canonical'' TL model is
\begin{equation}
v(R)=-1.038\left(\frac{GM}{R}\right)^{1/2}+1.196H_0R\, .
\label{velocity2}
\end{equation}
Notice that for v$(R_0)$=0, eq.~\ref{mass2} with the adequate numerical coefficient is recovered. In figure 3,
the $v$-$R$ relation in dimensionless variables is shown either for the ``canonical'' LT model (eq.~\ref{velocity2})
or the modified LT model (eq.~\ref{velocity1}). For radial distances smaller than $R_0$, the modified LT 
model gives
higher negative velocites when compared to the ``canonical'' LT model, consequence of an earlier turnaround.
For distances larger than $y \sim 1.39$ the acceleration becomes positive and velocities are slightly higher for
the modified LT model, reducing the distance within which the outer shells are gravitationally bound and, consequently
reducing the distance where the expanding shells merge with the Hubble flow.

\begin{figure}
\rotatebox{0}{\includegraphics[width=\columnwidth]{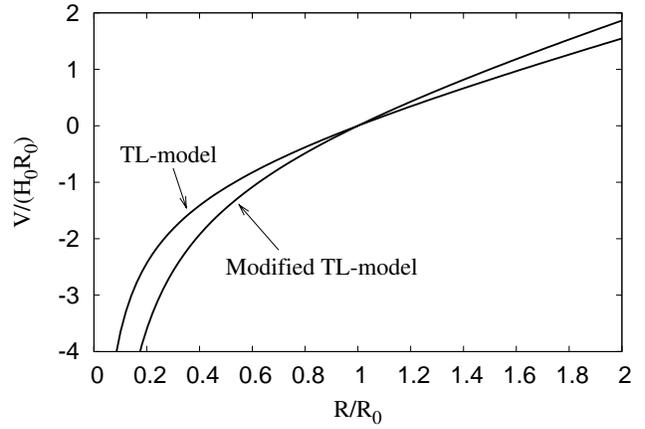}}
\caption{ Comparison between the $v$-$R$ relation  (dimensionless
 variables)  derived from the ``canonical'' LT model
 (eq.~\ref{velocity2}) and from the modified LT model (eq.~\ref{velocity1}).}
 \label{fig1}
 \end{figure}

The first point to be emphasized is that the fit quality (measured by the $\chi^2$ value) of both $v$-$R$ relations
(eqs.~\ref{velocity1}, \ref{velocity2}) to data is comparable. Thus, the resulting dispersion velocities of the
peculiar motion over the Hubble flow are practically the same for both models and are given in the last column
of table 3. Excluding the Virgo cluster, the mean value of the dispersion velocity for the other five groups is
$\sigma_{1D}=43\pm 7 km/s$, in agreement with the past estimates. 

Masses derived from the ``canonical'' LT model are given in the last column of table 2 and they differ, on
average, $\sim$ 10\% from the values estimated from the modified LT
model. It is worth mentioning that such
a difference is less than that expected by the use either of eq.~\ref{mass} or eq.~\ref{mass2}. In this
procedure the radius of the zero-velocity surface is determined independently and, as PP06 have shown,
in this case the resulting masses for the modified LT model are about 30\% higher than those derived from
the ``canonical'' model. By adopting the non-linear fit of the $v$-$R$ relation, the parameters are optimized
and the resulting $R_0$ value is not the same for both models but the masses are comparable. Such a method
also permits an ``optimized'' estimate of the Hubble parameter, given respectively in columns two (modified
LT model) and three (``canonical'' LT model) of table 3. Inspection of these figures reveals that the
Hubble parameter resulting from the fit of eq.~\ref{velocity2} is systematically higher by about 33\% than
those derived from the fit of eq.~\ref{velocity1}.

\begin{table}
\caption[]{The Hubble parameter derived from local flows: the second
column corresponds to values derived from the modified TL model, while the third
column corresponds to the ``canonical'' TL model. The last column gives the velocity
dispersion resulting from the fit of data to the $v$-$R$ relation for the modified TL model. The
mean value of the velocity dispersion excludes the Virgo cluster.}
\begin{flushleft}
\begin{tabular}{llll}
\hline
Group&h $(\Omega_{\Lambda}=0.7)$& h (TL model)&$\sigma$ (km/s)\\
\hline
M31/MW &$0.73\pm 0.04$&$0.87\pm 0.05$&38\\
M81 &$0.67\pm 0.04$&$0.82\pm 0.05$&53\\
NGC 253&$0.63\pm 0.06$&$0.74\pm 0.08$&45\\
IC 342&$0.57\pm 0.10$&$0.68\pm 0.12$&34\\
CenA/M83&$0.57\pm 0.04$&$0.68\pm 0.04$&45\\
Virgo&$0.71\pm 0.09$&$0.92\pm 0.12$&345\\
\hline
mean&$0.65\pm 0.07$&$0.79\pm 0.10$&$43\pm 7$\\ 
\hline
\end{tabular}
\end{flushleft}
\end{table}
 
Analysis of 3-year data of WMAP (Spergel et al. 2007) gives for the Hubble parameter $h=0.73\pm0.03$. 
However, studies of the local expansion flow lead to smaller values. Karachentsev et al. (2006) from the
analysis of 25 galaxies with velocities less than 500 km/s derived $h=0.68\pm 0.15$ and Sandage et al. (2006)
from a recalibration of distance indicators obtained $h=0.62\pm 0.05$. The resulting mean values given
in table 3 for both models are consistent with these determinations within the estimated uncertainties,
although the mean value of $H_0$ derived from the ``canonical'' TL model leads to an age for
the universe of $T_0\simeq 12.2$ Gyr, which seems to be a little short. 

\section{Conclusions}

The velocity profile for the LT model, modified by the inclusion of a cosmological constant, was calculated.
The inclusion of a such a term in the equation of motion modifies some characteristics 
of the ``canonical'' LT model. Shells inside the zero-velocity surface collapse earlier 
and, as a consequence, for a given distance negative velocities higher than those derived
from the ``canonical'' model are obtained. Moreover, shells
whose maximum expansion radius is beyond the critical value $R\sim 1.39R_0$ will never collapse, since
their acceleration becomes positive.

Data on dwarf galaxies belonging to nearby groups and galaxies in the outskirt of the Virgo cluster
are well represented by such a model, indicating that these objects are either collapsing or expanding,
to fallback in the future, if their distances are smaller than $\sim 1.39R_0$.
Moreover, such a good agreement between the theoretical $v$-$R$ relation and data
implies also that the use of the virial to estimate the core masses is questionable.

However, the $v$-$R$ relation derived from the ``canonical'' LT model gives an equally acceptable fit and
the present results cannot be used as an argument in favor of the detection of effects due to the
cosmological constant in scales of the order of few Mpc. Core masses derived from both models
agree to within 10\% but the Hubble parameter estimated from the ``canonical'' LT model is systematically 
higher than values resulting from the modified model. Nevertheless, mean values are compatible
with other independent estimates. 

The mean value of the 1-D dispersion velocity derived from the five groups (Virgo excluded) investigated
in this work is $43\pm 7 km/s$, smaller than  values derived from cosmological simulations
within scales of 1-3 Mpc, i.e., 73 km/s (PP06) and 80 km/s (Macci\`o et al. 2005). It is worth
mentioning that Axenides \& Perivolaropoulos (2002) studied the dark energy effects in the growth of 
matter fluctuations in a flat universe. They concluded that the dark energy can indeed cool
the local Hubble flow but the required parameters to make the predicted dispersion velocity
of the order of 40 km/s are ruled out by observations which constrain either the present
dark energy density or the equation of state parameter w(=$P_x/\varepsilon_x$). Thus,
the dark energy with a time independent equation of state cannot explain the observed quietness
of the local Hubble flow, which remains an enigma.

\begin{acknowledgements}

S. P. acknowledges the financial support through a ANR grant. We thank
 the referee for his useful comments which have contributed to improve
 the text of this paper.
\end{acknowledgements}

\end{document}